\documentclass[english,aps,floats,twocolumn,showpacs,nofootinbib]{revtex4}
\usepackage{pslatex}
\usepackage[T1]{fontenc}
\usepackage[latin1]{inputenc}
\usepackage{graphicx}
\usepackage{epsfig}
{\newcommand{\lsim}{\mbox{\raisebox{-.6ex}{~$\stackrel{<}{\sim}$~}}}
{ 
\newcommand{\be}{\begin{equation}}
\newcommand{\ee}{\end{equation}}
\newcommand{\bea}{\begin{eqnarray}}
\newcommand{\eea}{\end{eqnarray}}

\begin{document}
\title{Type-II Seesaw at Collider, Lepton Asymmetry and Singlet Scalar Dark Matter}
\author{John McDonald}
\email{j.mcdonald@lancaster.ac.uk}
\author{Narendra Sahu}
\email{n.sahu@lancaster.ac.uk}
\affiliation{Cosmology and Astroparticle Physics Group, University of 
Lancaster, Lancaster LA1 4YB, UK}
\author{Utpal Sarkar}
\email{utpal@prl.res.in}
\affiliation{Theory Division, Physical Research Laboratory,
Navarangpura, Ahmedabad, 380 009, India}
\begin{abstract}
We propose an extension of the standard model with a B-L global 
symmetry that is broken softly at the TeV scale. The neutrinos acquire 
masses through a type-II seesaw while the lepton (L) asymmetry arises in 
the {\it singlet sector} but without B-L violation. The model has the 
virtue that the scale of L-number violation ($\Lambda$) giving rise 
to neutrino masses is independent of the scale of leptogenesis 
($\Lambda'$). As a result the model can explain {\it neutrino 
masses, singlet scalar dark matter and leptogenesis at the TeV scale}. The 
stability of the dark matter is ensured by a surviving $Z_2$ symmetry, which could 
be lifted at the Planck scale and thereby allowing Planck scale-suppressed decay 
of singlet scalar dark matter particles of mass $\approx 3$ MeV to $e^+ e^-$ pairs 
in the Galactic halo. The model also predicts a few hundred GeV doubly charged scalar 
and a long lived charged fermion, whose decay can be studied at Large Hadron Collider 
(LHC) and International Linear Collider (ILC). 

\end{abstract}
\maketitle
\section{Introduction} 
Within the standard model (SM), neutrinos are massless. On the other 
hand, the current low energy neutrino oscillation data~\cite{solar_data,
atmos_data,kamland_data} indicate that at least two of the physical
left-handed (LH) neutrinos have tiny masses and therefore mix among 
themselves. However, as yet we do not know if the neutrinos are Dirac 
or Majorana. If the neutrinos are assumed to be Majorana then the sub-eV 
neutrino masses can be generated through the dimension five 
operator~\cite{weinberg.79}
\be
\mathcal{O}_\nu=\frac{\phi \phi \ell \ell}{\Lambda}\,,
\label{dim-5-operator}
\ee
where $\Lambda$ is the scale of lepton (L) number violation ($\Delta L=2$). 
The dimension five operator (\ref{dim-5-operator}) can originate through 
the celebrated see-saw mechanism. 

In type-I seesaw models, three right-handed (RH) neutrinos ($N$'s) are 
added without extending the gauge group of the SM. 
The canonical seesaw (or type-I seesaw)~\cite{canonical_seesaw} then 
gives the light neutrino mass matrix:
\be
m_\nu=m_\nu^I=-m_D M_N^{-1} m_D^T\,,
\label{typeI-seesaw}
\ee
where $m_D$ is the Dirac mass matrix of the neutrinos connecting
the LH neutrinos ($\nu_L$) with the RH neutrinos and $M_N$ is the 
Majorana mass matrix of the RH heavy neutrinos, which also sets the scale of 
L-number violation ($\Lambda$). The Dirac mass terms determine the L-numbers of 
the RH neutrinos to be $+1$ and hence the Majorana mass of the RH neutrinos 
violates L-number by two units. The decays of the RH neutrinos would then violate 
L-number and their CP violating out-of-equilibrium decay to SM fields can be a 
natural source of L-asymmetry~\cite{fukugita.86} in the early Universe. The CP-violation, 
which comes from the Yukawa couplings that determine the Dirac mass mass of neutrinos 
via the one-loop radiative vertex correction, requires at least two RH neutrinos. The 
masses of the RH neutrinos producing the final L-asymmetry then satisfy~\cite{type_I_group}
~\footnote{This could be next-to-lightest RH neutrino if flavor leptogenesis is 
considered~\cite{N2_leptogenesis}.}. 
\be
M_N\geq O(10^{9}) GeV\,.
\ee
If the corresponding theory of matter is supersymmetric (SUSY) then this bound, 
being dangerously close to the maximum reheat temperature, poses a problem. A modest 
solution was proposed in ref.~\cite{ma&sahu.06} by introducing an extra 
singlet heavy fermion. However, the model only achieves a reduction 
of above bound~\cite{type_I_group} by an order of magnitude.

In the type-II seesaw models, on the other hand, $SU(2)_L$ triplet Higgses ($\Delta$'s) 
are added to the SM gauge group. Explicit breaking of the L-number by trilinear 
couplings of the triplet Higgs scalar then induces a tiny vacuum expectation value (VEV) 
of the heavy triplet Higgs scalars~\cite{triplet_seesaw}, generating a light neutrino 
mass matrix:
\be
m_\nu=m_\nu^{II}=f\mu \frac{v^2}{M_{\Delta}^2}\,,
\label{typeII-seesaw}
\ee
where $M_{\Delta}$ is the mass of the triplet Higgs scalar $\Delta$,
$\mu$ is the coupling constant with mass dimension 1 for the trilinear term 
with the triplet Higgs and two standard model Higgs doublets, and $f$ is the 
Yukawa coupling of the triplet Higgs to the light leptons. 
$M_{\Delta}$ and $\mu$ are of the same order of magnitude and set
the scale ($\Lambda$) of L-number violation. $v$ is the VEV of the 
SM Higgs doublet. In these models, the L-asymmetry is generated through 
the L-number violating decays of the $\Delta$ to SM leptons and 
Higgs~\cite{ma&sarkar_prl}. The CP-violation, originating from the one-loop 
self-energy correction, requires at least two triplets. The scale of L-number 
violation is determined by $M_{\Delta}$ and $\mu$ and is required to be very 
high. In most cases the scale of L-violation in type-II seesaw models is larger than 
the type-I seesaw models~\cite{type_II_group}.

From the above discussion we see that the L-number violating scales $M_N\sim  
\Lambda$ in type-I see-saw models or $M_\Delta\sim \Lambda$ in type-II seesaw models 
comes out to be large because the same L-number violation gives rise to both neutrino 
masses and mixings and to the L-asymmetry. While sub-eV neutrino masses require large 
values of $\Lambda$, the large values of $M_N$ and $M_\Delta$ can easily satisfy the 
out-of-equilibrium decay condition, a necessary condition for the generation 
an L-asymmetry, without any fine-tuning of the Yukawa couplings. From 
this perspective, these models are attractive. However, they cannot be 
verified in future colliders since the scale of L-number violation is very 
high. Alternatives to these models are provided by mechanisms which
work at the TeV scale, either in SUSY extensions of the SM~\cite{susy_tev} or by introducing an additional
source of CP violation into the model~\cite{extra_tev}.

On the other hand, one could ask if the origin of neutrino masses and leptogenesis 
is different? It is well-known that there is no one-to-one correspondence 
between the parameters in the neutrino mass matrix and those involved in leptogenesis~\cite{
leptogenesis-mnu-3N}. In particular, in type-I see-saw model with 3 RH 
neutrinos, while 15 parameters enter into leptogenesis, there are only 
9 parameters: 3 masses, 3 mixing angles and 3 phases (one L-number conserving phase 
called the Dirac phase and two L-number violating phases called Majorana phases) in 
the low energy neutrino mass matrix. Obviously there is no connection. Conservatively, 
if one considers type-I seesaw model with 2 RH neutrinos for leptogenesis as well as 
for low energy sub-eV neutrino masses, then the number of parameters in both cases are 9. 
However, one can show that there is no one-to-one correspondence between the Majorana 
phases responsible for L-number violation in leptogenesis and Majorana phases in neutrino 
mass matrix~\cite{leptogenesis-mnu-2N}. The connection between leptogenesis and low energy 
neutrino mass matrix is even worse in case of type-II see-saw models.

Motivated by the fact that there is no one-to-one correspondence between 
leptogenesis and the effective low energy neutrino mass matrix, we propose a new 
mechanism of leptogenesis which occurs completely in the {\it singlet sector} at 
the TeV scale~\cite{singlet_lep}. As we will show, the origin of neutrino masses 
is different from the origin of L-asymmetry. The L-asymmetry then arises without any 
B-L violation. We will show that the B-L violation required for 
neutrino masses does not conflict with the leptogenesis. This model is then 
extended to incorporate a singlet scalar dark matter particle. While the nature of dark matter is still a 
mystery, we will show that the singlet dark 
scalar can either be collisionless cold dark matter (CCDM) or self-interacting dark 
matter (SIDM). The stability of the dark matter is ensured by a surviving $Z_2$ 
symmetry which could be lifted at the Planck scale and thereby allow Planck scale-suppressed decay of the singlet dark scalars to $e^+e^-$ pairs in the Galactic halo. The 
most important feature of the model is that it predicts a few hundred GeV doubly charged 
scalar and a long lived singly charged fermion whose decay can be studied at the LHC/ILC. 

The paper is organized as follows. In section II we introduce a model which simultaneously 
explains neutrino masses, singlet scalar dark matter and singlet leptogenesis. In section III, 
the type-II seesaw model of neutrino masses and the viability of testing it at colliders are discussed. 
In section IV we give a brief description of singlet leptogenesis, which arises from a 
conserved B-L symmetry. Section V is devoted to singlet scalar dark matter and its Planck 
scale suppressed decay to $e^+e^-$ pair in the Galactic halo. In section VI we give a brief 
description of collider signatures and section VII concludes.

\section{The Model: ${\rm SM}\times U(1)_{B-L}$} 
We extend the SM gauge symmetry with a global $U(1)_{B-L}$ symmetry, which 
is softly broken at the TeV scale. In addition to the quarks, leptons ($\ell$) 
and the usual Higgs doublet $\phi$ of the SM, we introduce two triplet scalars 
$\xi$ and $\Delta$ and a singlet complex scalar $\chi$. We also introduce 
two neutral singlet heavy scalars $S_a, a=1,2$ and a charged fermion $\eta^-$. 
The particle content of the model and their respective quantum numbers is given 
in table (\ref{parti}).
\begin{table}[h]
\begin{center}
\caption{Fermions and scalars included in the model.}
\label{parti}
\begin{tabular}{|c|c|c|}\hline
Particle Content & $SU(3)_C\times SU(2)_L\times U(1)_Y$ &  $U(1)_{B-L}$\\ \hline
$\ell$  & (1,2,-1) & -1 \\[2mm] \hline
$e_R^-$ & (1,1,-2) & -1 \\[2mm] \hline
$\phi$  & (1,2,1) & 0\\[2mm] \hline
$\xi$   & (1,3,2) & 2 \\[2mm] \hline
$\Delta$ & (1,3,2) & 0\\[2mm] \hline
$\chi$ & (1,1,0) & 0\\[2mm] \hline
$\eta_L^-, \eta_R^-$ & (1,1,-2) & -1\\[2mm] \hline
$S_a$    & (1,1,0)  & 0 \\[2mm] \hline
\end{tabular}
\end{center}
\end{table}

It is then straightforward to write down the Lagrangian invariant under the 
SM and the global $U(1)_{B-L}$. We present here only those terms in the
Lagrangian that are directly relevant to the rest of our discussions. Those 
are given by
\bea
-{\cal L } &\supseteq & f_{ij} \xi \ell_{iL} \ell_{jL} + \mu \Delta^\dagger 
\phi \phi + M_\xi^2  \xi^\dagger \xi + M_\Delta^2 \Delta^\dagger 
\Delta \nonumber\\
&&+ h_{ia} \bar e_{iR} S_a \eta_L^- + M_{S_{ab}}^2 S_a^\dagger S_b + y_{ij}\phi
\bar \ell_{iL} e_{jR} \nonumber\\
&& +g_{i1} \bar \ell_{iL}\phi \eta_R^- + M_\eta \overline{\eta_L^-} \eta_R^-
+\mu_{\phi a} S_a \phi^\dagger \phi \nonumber\\ 
&&+\mu_{\chi a} S_a \chi^\dagger \chi + V({\phi, \chi, S, \Delta}) + h.c.\,,
\label{lagrangian}
\eea
where $V (\phi, \chi,S, \Delta)$ constitutes all possible quadratic and 
quartic terms invariant under the SM and the global $U(1)_{B-L}$,
\bea
V(\phi, \chi, S, \Delta)&=& m_1^2 \phi^\dagger \phi +m_2^2 \chi^\dagger 
\chi+\lambda_1 (\phi^\dagger \phi)^2+\lambda_2 (\chi^\dagger\chi)^2\nonumber\\
+\lambda_3(\phi^\dagger \phi)(\chi^\dagger \chi) &+& \lambda_a(S_a^\dagger S_a)
(\phi^\dagger \phi) + \lambda_b(S_b^\dagger S_b)(\chi^\dagger \chi)\nonumber\\
+\lambda_4 (\Delta^\dagger \Delta)(\phi^\dagger \phi) &+& \lambda_5 
(\Delta^\dagger \Delta) (\chi^\dagger \chi)\,.
\eea

As the Universe expands, $\Delta$ acquires a very small VEV, 
\begin{equation}
\langle \Delta \rangle = -\mu  {v^2 \over M_\Delta^2}\,.
\end{equation}
For $\mu\sim M_\Delta\sim 10^{12}$ GeV and $v= \langle \phi \rangle =174$ GeV one 
can get the VEV of $\Delta$ to be a few eV. The singlet scalars $S_1$ and $S_2$ 
acquire VEV much below the mass scale of $\Delta$. They develop VEV at 
a temperature $T\sim 10$ TeV. The VEVs of $S_1$ and $S_2$ are:
\be
\langle S_a \rangle =-\mu_{\phi a} \frac{v^2}{M_{S_a}^2}\,,\,\,\, a=1,2\,.
\ee
For $\mu_{\phi a}\sim M_{S_a}\sim 10$ TeV, the VEV of $S_a$ are in the order 
of 1 GeV. The singlet scalar $\chi$ does not acquire a VEV. With a $Z_2$ 
symmetry under which $\chi$ is odd while all other fields are even, $\chi$ 
becomes a candidate for dark matter~\cite{jm1,jm2,darkmatter}. As we will see below,
the L-asymmetry arises from the $L$-number conserving decay of $S_a$
to $e_R^-$ and $\eta_L^+$ while the neutrino masses arise from the 
$L$-violating interactions produced via a soft breaking interaction of 
$\xi$ and $\Delta$. 

\section{Soft Breaking of B-L Symmetry and Neutrino Masses}
The VEV of $\Delta$ does not break B-L gauge symmetry since it is 
inert under the $U(1)_{\rm B-L}$ symmetry. This also ensures that 
B-L is an exact symmetry until it is broken softly at the TeV scale. 
Note that $\xi$ does not acquire any VEV at the tree level and hence 
there are no neutrino masses unless the B-L symmetry is broken. 
We assume that $M_\xi \ll M_\Delta$ and that $\xi$ and $\Delta$ 
contribute equally to the effective neutrino masses.

To generate neutrino masses we need to break the global $U(1)_{\rm B-L}$ symmetry
without destroying the renormalizability of the theory while ensuring
that there is no massless Nambu-Goldstone boson that can cause conflict
with phenomenology. This can be achieved by adding a soft term
in the Lagrangian (\ref{lagrangian})
\begin{equation}
-{\cal L }_{\Delta \xi} = m_{s}^2 \Delta^\dagger \xi + h.c.\,,
\label{soft_term}
\end{equation}
where the mass parameter $m_s$ is of the order of a few hundred GeV. We assume that 
such a soft term could originate from theories with larger symmetries. However, we will not consider its origin in this paper. The mixing between $\xi$ and $\Delta$ is then parameterized by
\be
\tan 2 \theta=\frac{2 m_s^2}{M_\Delta^2 - M_\xi^2}
\ee
Since we have assumed that $M_\Delta\gg M_\xi$, the mixing angle is simply 
\be
\theta \simeq \frac{m_s^2}{M_\Delta^2}\,.
\ee
As a result the mass eigenstates are:
\be
\xi'=\xi-\left( \frac{m_s^2}{M_\Delta^2} \right) \Delta \simeq \xi \; {\rm and} \; \Delta' =
\Delta + \left( \frac{m_s^2}{M_\Delta^2} \right)\xi \simeq \Delta\,.
\ee
Since the soft term (\ref{soft_term}) introduces L-number violation by two units, the neutrino
can acquire a mass. The effective L-number violating Lagrangian is~\footnote{We 
thank Hiroaki Sugiyama for pointing out a typographical mistake in equation (\ref{flavour_vio}).}:
\begin{equation}
-{\cal L }_{\nu - mass} = f_{ij} \xi \ell_{iL} \ell_{jL} + \mu
{m_s^2 \over M_\Delta^2} \xi^\dagger \phi \phi + f_{ij} {m_s^2 \over M_\Delta^2}
\Delta \ell_{iL} \ell_{jL} + \mu \Delta^\dagger \phi \phi + h.c.\,.
\label{flavour_vio}
\end{equation}

\begin{figure}
\begin{center}
\epsfig{file=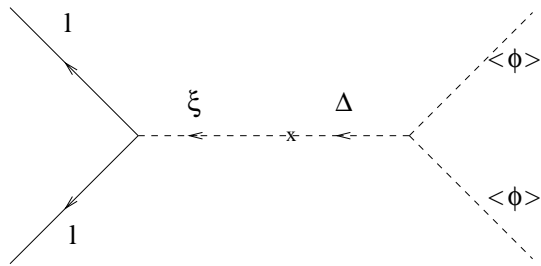, width=0.4\textwidth}
\caption{Modified type-II seesaw for neutrino masses arising from soft L-number
violation. \label{seesaw}}
\end{center}
\end{figure}

After electroweak symmetry breaking, the field $\xi$ acquires an induced VEV,
\be
\langle \xi \rangle = -\mu {v^2 m_s^2 \over M_\xi^2 M_\Delta^2}\,.
\label{xi_vev}
\ee
This can be verified by minimization of the complete potential. The VEVs of $\xi$ and $\Delta$ will contribute equally to neutrino masses and thus 
the neutrino mass matrix, derived from fig. (\ref{seesaw}), is given by
\begin{equation}
\left( {m}_{\nu} \right)_{ij} = - f_{ij} \mu {v^2 m_s^2 \over M_\xi^2 M_\Delta^2}\,.
\label{neutrino_mass}
\end{equation}
If we consider the mass scales $\mu\sim M_\Delta\sim 10^{12}$ GeV,
$m_s\sim 100$ GeV and $M_\xi \sim v$, a natural choice of the
Majorana Yukawa coupling $f$ gives the scale of neutrino masses to be 
$m_\nu \sim {\mathcal O}(1)$ eV, as required by laboratory, 
atmospheric and solar neutrino experiments. As discussed previously, one 
of the triplet Higgs scalar $\xi$ could remain very light without any conflict
with neutrino masses or any other phenomenology. Since the mass of $\xi$
could be in the range of a few hundred GeV, its decay through same sign 
dilepton can be tested at the LHC or ILC. Thus the proposed type-II seesaw is 
testable in contrast to the conventional type-II seesaw. We will come back 
to this point in section VI while discussing collider signatures.

\section{Lepton asymmetry from conserved B-L}
We note that the interaction $S_a\eta_L^- e_R^+$ conserves B-L-number. Therefore, 
out-of-equilibrium decay of $S_a$ cannot generate any B-L asymmetry. 
However, if there is CP-violation in the decay of $S_a$ then it can 
produce an equal and opposite B-L asymmetry between $e_R^+ (e_R^-)$ and 
$\eta_L^- (\eta_L^+)$. If these two B-L asymmetries never equilibrate with each other 
before the electroweak phase transition then the B-L asymmetry in 
$e_R^-$ can be transferred to the left-handed fields in SM, while keeping an 
equal and opposite B-L asymmetry in $\eta_L^+$. The Yukawa couplings of the
charged leptons with the SM Higgs field will transfer any asymmetry in
$e_R^-$ into an asymmetry in $e_L^-$, which will then take part in 
sphaleron processes before the electroweak phase transition. Thus the
B-L asymmetry in $e_R^-$ will generate the required baryon asymmetry
via sphaleron transitions, while the equal and opposite amount of
B-L asymmetry in $\eta_L^+$ will remain unaffected even after the
electroweak phase transition. Eventually this field will decay slowly
after the electroweak phase transition, which can generate a small
L-asymmetry, but the baryon asymmetry of the universe will not
be affected at this time since the sphaleron processes are out of thermal 
equilibrium. Therefore the baryon asymmetry will survive.

\begin{figure}[htbp]
\begin{center}
\epsfig{file=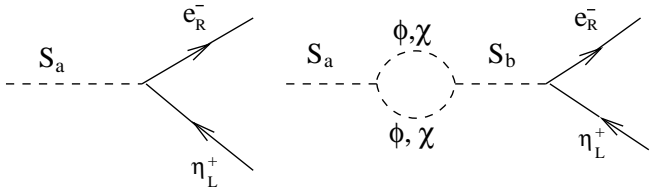, width=0.48\textwidth}
\caption{The tree level diagram and the self energy correction diagram 
of $S_a$ which give a net CP violation.}
\label{fig1}
\end{center}
\end{figure}

As the Universe expands, the temperature of the thermal bath falls. Below 
their mass scales the CP violating decay of the heavy singlet scalars 
$S_a, a=1,2$ generate an equal and opposite B-L-asymmetry between 
$e_R^-(e_R^+)$ and $\eta_L^+(\eta_L^-)$ through 
\begin{eqnarray}
S_a &\to & e_{iR}^- + \eta_L^+ \nonumber \\
& \to & e_{iR}^+ + \eta_L^- \,.\nonumber
\end{eqnarray}
The decay rate can be given as 
\begin{equation}
\Gamma_a=\frac{(h^\dagger h)_{aa}}{16 \pi} M_{S_a}\,,
\end{equation}
where the masses of $\eta_L^-$ and $e_R^-$ are small in comparison to 
the mass of $S_a$. A net CP asymmetry is generated through the interference 
of tree-level diagram and the one-loop self-energy correction diagram involving 
$\phi$ and $\chi$ as shown in figure (\ref{fig1}). 
The CP-asymmetry in the decay of $S_a$ can be estimated as
\begin{equation}
\epsilon_a=\frac{ {\mathrm Im}\left[ (\mu_{\phi 1}\mu_{\phi 2}^*+
\mu_{\chi 1}\mu_{\chi 2}^*) \sum_i h_{i1} h_{i2}^* \right]} 
{8 \pi^2 (M_{S_1}^2-M_{S_2}^2)} \left[ \frac{M_{S_a}}{\Gamma_a}\right]\,.
\end{equation} 
The lightest of $S_a$ will generate an equal amount of $e_R$ and $\eta$
asymmetries. If the masses of $S_1$ and $S_2$ are close enough then the CP 
asymmetry can be resonantly enhanced,~\cite{singlet_resonant,triplet_resonant} 
such that the mass scale of $S_a$ can be a few TeV. 

\begin{figure}[htbp]
\begin{center}
\epsfig{file=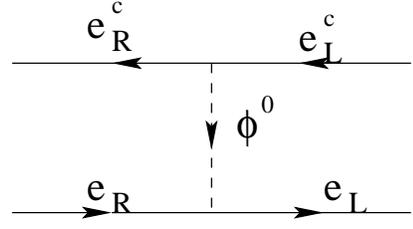, width=0.3\textwidth}
\caption{The L-number conserving process which transfer the B-L 
asymmetry from right handed sector to the left-handed sector.}
\label{fig2}
\end{center}
\end{figure}

The L-asymmetry in $e_R$ can be transferred to $e_L$ through the 
$t$-channel process $e_R e_R^c \leftrightarrow \phi_0 \leftrightarrow 
e_L e_L^c$ as shown in the figure (\ref{fig2}).
These interactions will be in equilibrium for all the three
generations of charged leptons below $10^5$~GeV and hence there will
be equal amount of $e_R$ and $e_L$ asymmetry. This B-L asymmetry
in $e_L$ will be converted to the baryon asymmetry of the universe
before the electroweak phase transition when the sphaleron processes are
in thermal equilibrium. An equal and opposite amount of B-L
asymmetry in $\eta_L$ will remain unaffected by these interactions.

\begin{figure}
\begin{center}
\epsfig{file=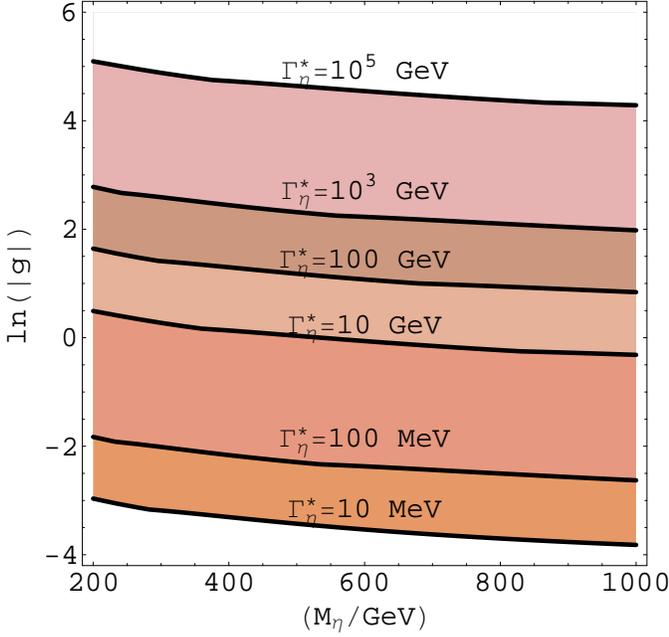, width=0.5\textwidth}
\caption{Allowed contours of $\Gamma^*_\eta\equiv 10^{19}\Gamma_\eta$, required for 
generating a successful L-asymmetry, are shown in the plane of $\ln |g|$ versus 
$M_\eta$.\label{eta-contour}}
\end{center}
\end{figure}

Note that the generated B-L asymmetry in the left-handed sector is 
not washed out by the L-violating interactions mediated by $\xi$ and 
$\Delta$ because those processes are suppressed. In particular $\ell 
\ell\leftrightarrow \phi \phi$ is suppressed by $m_s^2/M_\Delta^2$. For 
$m_s\sim 1$ 100 GeV and $M_\Delta\sim 10^{12}$ GeV the suppression of 
$\Delta L=2$ processes are of the order $10^{-20}$. However, this 
asymmetry can be washed out through the decay: $\eta_R^+ \to \bar{\ell} + 
\phi$, unless the decay rate
\be 
\Gamma_\eta=\frac{|g_{i1}|^2}{16 \pi} M_\eta\,,
\label{eta_decay}
\ee
satisfies $\Gamma_\eta^{-1} \equiv \tau_\eta > \tau_{EW}$, where $\tau_{EW}\sim 
10^{-12}$s is the time of electroweak phase transition. 
Furthermore, $\eta^+$ should be decayed away well before Big-Bang 
Nucleosynthesis (BBN) in order not to conflict with the prediction of 
BBN. Therefore
\be
\tau_\eta < \tau_{BB} \sim 1s\,.
\label{eta_bigbang}
\ee
From Eqns. (\ref{eta_decay}) and (\ref{eta_bigbang}), and using $\tau \approx H^{-1} 
\approx (T^2/M_{Pl})^{-1}$, we get 
\be
T_{BB}^{2}/M_{Pl} \lsim \frac{|g_{i1}|^2}{16 \pi} M_\eta \lsim T_{EW}^{2}/M_{Pl}\,,
\label{g-range}
\ee
where $T_{EW}$ and $T_{BB}$ respectively are the temperatures corresponding 
to the electroweak phase transition and BBN. We 
show the allowed masses and Yukawa couplings of $\eta$ as a plot of $\ln |g|$ versus $M_\eta$. From figure (\ref{eta-contour}) it can be 
seen that $g$ can vary from $10^{-7}$ to $10^{-12}$ for $M_\eta$ taking 
its values between 200 GeV to 1 TeV. 

\section{Singlet Scalar Dark Matter} 
As the universe expands the temperature of the thermal bath falls. As a result 
the heavy fields $S_a$ and $\Delta$ acquires VEV below their mass scales. Consequently 
the effective potential before the electroweak phase transition is given by 
\bea
V_{eff} &=& m_\phi^2 \phi^\dagger \phi +\lambda_1 (\phi^\dagger \phi)^2 
+m_\chi^2 \chi^\dagger\chi \nonumber\\
&+& \lambda_2 (\chi^\dagger\chi)^2 +\lambda_3(\phi^\dagger \phi)(\chi^\dagger \chi)
\eea
where 
\bea
m_\phi^2 &=& \left( m_1^2 + \lambda_a \langle S_a \rangle^2+ \lambda_4 \langle \Delta       
\rangle^2\right)\nonumber\\
{\rm and} 
~~~m_\chi^2 &=& \left( m_2^2 + \lambda_b \langle S_b \rangle^2 + \lambda_5 \langle \Delta
\rangle^2 \right)\,.
\eea
The above effective potential is bounded from below if and only if $\lambda_1, 
\lambda_2 >0$ and $\lambda_3>-2\sqrt{\lambda_1 \lambda_2}$. For $m_\phi^2 <0$ and 
$m_\chi^2>0$, the minimum of the potential is given by 
\be
\langle \phi \rangle =\pmatrix{ 0\cr v } ~~~~~{\rm and}~~~~~  
\langle \chi \rangle=0\,.
\ee
The VEV of $\phi$ gives masses to the SM fermions and gauge bosons. 
The physical mass of the SM Higgs is then given by $m_h = \sqrt{4\lambda_1 v^2}$. 
Since $\chi$ is odd under the surviving $Z_2$ symmetry it cannot decay 
to any of the conventional SM fields and hence the $\chi$ can constitute the 
dark matter component of the Universe. 

\subsection{Cold Dark Matter}

Gauge singlet scalars interacting via the renormalisable $\chi^{\dagger}\chi 
\phi^{\dagger}\phi$ coupling to the Higgs doublet can account for cold dark 
matter (CDM) \cite{jm1,jm2,darkmatter}\footnote{The notation used in \cite{jm1,jm2} is 
such that $S \leftrightarrow \chi$, $\lambda_{S} \leftrightarrow \lambda_{3}$, $m 
\leftrightarrow m_{\chi}$,  $m_{S} \leftrightarrow m_{\chi\;tot}$ and $\eta 
\leftrightarrow 4 \lambda_{2}$.}.  For $\lambda_{3} \; ^{>}_{\sim} \; 0.01$, 
the gauge singlet scalar CDM density is produced via conventional freeze-out 
from thermal equilibrium. For $ 10 {\rm GeV}  \; ^{<}_{\sim} \; m_{\chi\;tot}  
\; ^{<}_{\sim} \; 1 {\rm TeV}$ and $0.01 \; ^{<}_{\sim} \; \lambda_{3} \; ^{<}_{\sim} 
\; 1$, the scalar density is naturally of the order of the observed CDM density 
over a wide region of the $(\lambda_{3}, m_{\chi\;tot})$ parameter space \cite{jm1}. 
($m_{\chi\;tot}$ denotes the physical $\chi$ mass, including the contribution 
from the Higgs expectation value, $m_{\chi\;tot}^{2} = m_{\chi}^{2} + \lambda_{3} 
<\phi^{\dagger}\phi>$.)  Such scalars have annihilation and nuclear scattering 
cross-sections which make them a potentially detectable CDM candidate in 
cryogenic detectors and neutrino telescopes, on a par with more conventional 
WIMP candidates \cite{jm1}. They may also be produced at the LHC via Higgs 
decay \cite{jm1,darkmatter}.

For small $\lambda_{3}$ there is an alternative possibility for gauge singlet 
scalar dark matter \cite{jm2}. If $\lambda_{3} \ll 1$, gauge singlet scalars 
in thermal equilibrium are unable to annihilate efficiently, resulting in 
too much CDM after freeze-out \cite{bento1}. In order to evade this problem, 
$\lambda_{3}$ must be sufficiently small that the gauge singlets never come 
into thermal equilibrium. Although there is no freeze-out thermal relic 
density in this case, it is still possible to produce a CDM density via 
decay of thermal equilibrium Standard Model particles to $\chi$ pairs, in 
particular Higgs decay \cite{jm2}. In this case the density of CDM is given 
by \cite{jm2} 
\be \Omega_{\chi} =
0.3 \left( \frac{\lambda_{3}}{2 \times 10^{-10}} \right)^{2} 
\left( \frac{115 {\rm GeV}}{m_{h}} \right)^{3}
\left( \frac{0.7}{ h } \right)^{2}  
\left( \frac{m_{\chi\;tot}}{4.8 {\rm MeV}} \right)^{2}
  ~,\ee
where $m_{h}$ is the physical Higgs boson mass and $h$ parameterizes the Hubble constant. 
Thus $\lambda_{3} \approx 10^{-10}$ is necessary to account for CDM. 
A particularly interesting possibility is that $\chi$ gains its mass mostly from the 
Higgs expectation value. In general $m_{\chi\;tot}^2 = m_{\chi}^{2} + \lambda_{3} v^{2}/2$ 
($v  = 174$ GeV). Therefore for $m_{\chi} = 0$, the $\chi$ mass is a function only of 
$\lambda_{3}$. In this case the dark matter density is entirely determined by 
$m_{\chi\;tot}$ and $m_{h}$ \cite{jm2}, 
\be 
\Omega_{\chi} = 0.3 \left( \frac{115 {\rm GeV} }{m_{h}} \right)^{3}
\left( \frac{0.7}{ h } \right)^{2}  
\left( \frac{m_{\chi\;tot}}{2.8 {\rm MeV}} \right)^{5}
  ~.
\ee
Thus $m_{\chi\;tot} \approx 3$ MeV is predicted for the scalar mass in this case. 
(Note that this is a rather precise estimate, due to the high power of $m_{\chi\;tot}$ 
in Eq.(25).) As discussed below, this is a physically significant mass scale. Light 
gauge singlet scalars with masses in the 1-10 MeV range \cite{bento2,jm2} can account 
for self-interacting dark matter (SIDM) \cite{sidm}, whilst the decay of 
1-6 MeV scalars to $e^{+}$ $e^{-}$ pairs can account for the 511 keV $\gamma$ line 
from the galactic center \cite{boehm,posp,beacom} observed by INTEGRAL \cite{integral}. 

\subsection{Self-Interacting Dark Matter}

SIDM has been suggested as a way to explain the absence of peaked galaxy halo profiles 
and the small number of sub-halos in galaxies as compared with CDM simulations \cite{sidm}. 
This requires that the self-interaction cross-section of the dark matter, $\sigma$, 
satisfies $ r_{\chi} \equiv \sigma/m_{\chi} = (0.45 - 5.6) {\rm cm^{2} g^{-1}}$ \cite{sidm}. 
Recently, comparison of the theoretical dynamics of the merging galaxy cluster 1E 0657-56 
(the 'Bullet cluster') with observation has placed a new upper bound on this range, 
$r_{\chi} \;  ^{<}_{\sim} \; 1 {\rm cm^{2} g^{-1}}$ \cite{mark}. Nevertheless, a window 
still remains where SIDM scalars could have a significant effect on galaxy halos.

For a gauge singlet self-interaction $\lambda_{2} |\chi|^{4}$, the total scattering 
cross-section is \cite{jm2} 
\be 
\sigma = \sigma_{\chi \chi^{\dagger} \rightarrow \chi \chi^{\dagger}} + 
\sigma_{\chi \chi \rightarrow \chi \chi}
   = \frac{3 \lambda_{2}^{2}}{8 \pi m_{\chi}^2}    ~. 
\ee
This self-interaction can modify galaxy halos if the mass is in the range 
$m_{S} = \alpha_{\lambda_{2}}^{1/3} \left(38.8-90.1\right)$  MeV (with $\alpha_{\lambda_{2}} 
= \lambda_{2}^{2}/4 \pi$), where the lower bound is the value at which galaxy halos would 
evaporate due to interaction with hot particles in cluster halos, while the upper bound is 
the value at which the scalars do not interact within a typical galactic halo during a 
Hubble time and so have no effect \cite{sidm,jm2}. (The new constraint from the Bullet 
cluster raises the lower bound to $68.7 \alpha_{\lambda_{2}}^{1/3}$ MeV.)  If we consider 
the magnitude of the $\chi$ self-coupling to be similar to the SM Higgs self-coupling, such 
that $\lambda_{2} \approx 0.025$, then this mass range is $m_{\chi} \sim 1-4$ MeV. In this 
case there is a good possibility that light scalar dark matter will have an observable 
effect on galaxy halos.

\subsection{Decaying $\chi$ Dark Matter and the INTEGRAL 511 keV flux}

In order to explain the 511 keV $\gamma$-ray flux observed by INTEGRAL, we need a 
large number density of dark matter particles at the center of the galaxy plus 
low energy positrons from their decay in order that the positrons can slow to 
non-relativistic velocities before annihilating with background electrons. These 
conditions require that the decaying particles are in the range 1-6 MeV, corresponding to positron injection energies $\lsim 3$ MeV \cite{beacom}, in good agreement with $m_{\chi} \approx 3$ MeV. 
For an interaction of the form  $ \chi m_{e} \overline{e} e/M_{*}$, the $\gamma$ flux 
relative to the experimentally observed flux is \cite{posp}  
\be 
\frac{\Phi}{\Phi_{exp}} \approx 25 \left( \frac{10^{19} {\rm GeV}}{ M_{*} } 
\right)^{2} \left( \frac{\Omega_{\chi}}{0.23} \right) ~,
\ee
where the dominant decay mode is assumed to be $\chi \rightarrow e^{+}e^{-}$ with 
decay rate 
\be 
\Gamma_{\chi \rightarrow e^{+}e^{-}} = \frac{m_{e}^{2} m_{\chi}}{8 \pi M_{*}^{2}} ~.
\ee  
Thus with a Planck-suppressed decay rate, light gauge singlet scalars of mass 
$m_{\chi} \approx 3$ MeV can account for the observed 511 keV flux. 

The above decay: $\chi \rightarrow e^{+}e^{-}$ can be realized if we allow 
$Z_2$ to be violated at the Planck scale. We have then the allowed Planck 
scale suppressed Lagrangian
\bea
\mathcal{L} &=& y\frac{\chi\phi \bar{\ell}e_R}{M_{\rm Pl}}+f \frac{\chi \xi 
\ell \ell}{M_{\rm Pl}}+f \frac{\chi (m_s^2/M_\xi^2)\Delta \ell \ell}
{M_{\rm Pl}}\nonumber\\
&&+g \frac{\chi \bar{\ell}\phi \eta_R^-}{M_{\rm Pl}}+
h\frac{\chi S \eta_L^- e_R^+}{M_{\rm Pl}}+\hat{h}\chi \eta_L^- e_R^+ + h.c. \,.
\eea
Since $m_\chi \ll M_\eta$, the decay modes: $\chi \rightarrow \eta_L^- e_R^+$ 
and $\chi \rightarrow \bar{\ell} \phi \eta_R^-$, $S \eta_L^- e_R^+$ are forbidden. 
Thus the relevant effective $Z_2$ violating Lagrangian is given as:
\be
\mathcal{L} = \frac{m_e}{M_{\rm Pl}}\chi \bar{e_L}e_R + f' \chi \ell \ell 
+f' (m_s^2/M_\xi^2)\chi \ell \ell + h.c. \,,
\ee
where $f' \equiv f' (\langle \Delta \rangle /M_{\rm Pl})\leq 10^{-26}$. 
Therefore, the contribution of the second and third terms are negligible 
to the observed $\gamma$-ray flux while the first term is in the right 
ballpark.

\section{Signatures of $\xi^{\pm \pm }$ and $\eta^\pm$} 
The doubly charged component of the light triplet Higgs $\xi$ can be 
observed through its decay into same sign dileptons~\cite{collider_signature}. 
Since $M_\Delta \gg M_\xi$, the production of $\Delta$ particles in 
comparison to $\xi$ is highly suppressed. Hence it is worth looking
for the signature of $\xi^{\pm\pm}$ either at LHC or ILC. 
From Eq. (\ref{flavour_vio}) one can see that the decay $\xi^{\pm \pm} 
\rightarrow \phi^\pm \phi^\pm$ are suppressed since the decay rate involves 
the factor $\frac{m_s^2}{M_\Delta^2}\sim 10^{-20}$. While the decay mode $\xi^{\pm\pm} 
\rightarrow h^\pm W^\pm $ is phase space suppressed, the decay mode $\xi^{\pm\pm} 
\rightarrow W^\pm W^\pm $ is suppressed because the VEV of $\xi$ is 
small as required for sub-eV neutrino masses and to maintain 
the $\rho$ parameter of SM to be unity. Therefore, once produced, 
$\xi$ mostly decays through same sign dileptons: $\xi^{\pm \pm} \rightarrow 
\ell^\pm \ell^\pm$. Note that the doubly charged particles cannot couple 
to quarks. Therefore the SM background of the process $\xi^{\pm\pm}
\rightarrow \ell^\pm \ell^\pm$ is quite clean and so the detection will 
be unmistakable. From Eq. (\ref{flavour_vio}) the decay rate of 
the process $\xi^{\pm\pm}\rightarrow \ell^\pm \ell^\pm$ is given by 
\be
\Gamma_{ii}=\frac{|f_{ii}|^2}{8\pi} M_{\xi^{++}} ~~~~{\mathrm and }~~~~ 
\Gamma_{ij}=\frac{|f_{ij}|^2}{4\pi} M_{\xi^{++}}\,, 
\ee
where the $f_{ij}$ are highly constrained by lepton flavor violating 
decays. Therefore if a doubly charged scalar can be detected in the future, the neutrino mass patterns can be probed at a collider~\cite{neutrinomass_pattern}~\footnote{There has been recent interest in detecting neutrino mass parameters at colliders~\cite{newinterests}. 
These references appeared on arXiv after our paper.}.  

Since the mass of $\eta^-$ is a few hundred GeV, the decay $\eta^-\rightarrow 
h e^-$ can be observed in future colliders (LHC/ILC). On the other hand, 
$\eta^-$ can also be produced in high energy neutrino collisions with matter 
through $\overline{\nu_L} e_R^- \rightarrow \overline{\nu_L} \eta_R^-$. Since 
it is long-lived, it can travel large distances and be detected in neutrino 
telescopes~\cite{eta_test}~\footnote{We thank John F. Beacom for bringing this to our 
notice.}.  

\section{Conclusions} 
We have introduced a new leptogenesis mechanism in a $U(1)_{B-L}$ extension of the 
SM which allowed us to explain simultaneously neutrino 
masses, dark matter and leptogenesis. The important message is that the L-asymmetry 
arises without any B-L violation. Since the L-number violation required for leptogenesis 
and neutrino masses are different, the leptogenesis scale can be lowered to as low as a 
few TeV.

Neutrino masses arise through a modified type-II seesaw which predicts a 
few hundred GeV triplet scalar. Note that in conventional type-II seesaw 
models the mass scale of the triplets is required to be at least
$O(10^{10})$ GeV to produce sub-eV neutrino masses. Since one of the triplets in 
our model, namely $\xi$, has a mass of a few hundred GeV, the proposed model can 
be tested in the near future at colliders through same sign dilepton decay of 
$\xi$. 

The model also predicts a singly charged fermion $\eta^-$ of mass ranging from 200 GeV 
to 1 TeV. $\eta^-$ can be produced in high energy neutrino collision with matter. 
Since it is long-lived, it can travel large distances and be observed in neutrino 
telescopes.

We proposed a singlet scalar, $\chi$, with mass $\approx 3$ MeV as a candidate for 
dark matter, whose stability is ensured by a $Z_2$ symmetry. This $Z_2$ discrete 
symmetry can be broken at the Planck scale. We then showed that a possible origin 
of 511 KeV Galactic line detected by INTEGRAL could be the Planck scale-suppressed 
decay of $\chi$ to $e^- e^+$ pairs in the Galactic halo. 

\section*{Acknowledgement} 

JM and NS were supported by the European Union through the Marie Curie 
Research and Training Network "UniverseNet" (MRTN-CT-2006-035863) and by 
STFC (PPARC) Grant PP/D000394/1.


\begin{thebibliography}{99}



\bibitem{solar_data} Q.R.~Ahmed {\it et al} (SNO Collaboration),
Phys.\ Rev.\ Lett. {\bf 89}, 011301-011302 (2002);
J.N.~Bahcall and C.~Pena-Garay, [arXiv:hep-ph/0404061].

\bibitem{atmos_data} S.~Fukuda {\it et al} (Super-Kamiokande
Collaboration), Phys.\ Rev.\ Lett. {\bf 86}, 5656 (2001).

\bibitem{kamland_data} K.~Eguchi {\it et al} (KamLAND collaboration),
Phys.~Rev.~Lett. {\bf 90}, 021802 (2003).

\bibitem{weinberg.79} S.~Weinberg,
  Phys.\ Rev.\ Lett.\  {\bf 43}, 1566 (1979).

\bibitem{canonical_seesaw} P. Minkowski, Phys. Lett. {\bf B 67}, 421 (1977);
M.~Gell-Mann, P.~Ramond and R.~Slansky in {\it Supergravity} (P.~van 
Niewenhuizen and D.~Freedman, eds), (Amsterdam), North Holland, 1979; 
T.~Yanagida in {\it Workshop on Unified Theory and Baryon number in the 
Universe} (O. Sawada and A.~Sugamoto, eds), (Japan), KEK 1979; 
R.N.~Mohapatra and G.~Senjanovic, Phys.\ Rev.\ Lett. {\bf 44}, 912 (1980)\,.

\bibitem{fukugita.86} M.~Fukugita and T.~Yanagida, Phys. Lett.
{\bf B174}, 45 (1986)\,.

\bibitem{type_I_group} S.~Davidson and A.~Ibarra,
Phy\ . Lett.\ B{\bf 535}, 25 (2002); W.~Buchmuller, P.~Di Bari and
M.~Plumacher, Nucl.\ Phys.\ B{\bf 643}, 367 (2002)\,.

\bibitem{N2_leptogenesis} See for example G.~Engelhard, Y.~Grossman, E.~Nardi and Y.~Nir,
  Phys.\ Rev.\ Lett.\  {\bf 99} (2007) 081802 \,.

\bibitem{ma&sahu.06} E.~Ma, N.~Sahu and U.~Sarkar,
  J.\ Phys.\ G {\bf 34}, 741 (2007)
  [arXiv:hep-ph/0611257]; 
  J.\ Phys.\ G {\bf 32}, L65 (2006)
  [arXiv:hep-ph/0603043].

\bibitem{triplet_seesaw} J. Schechter and J.W.F. Valle, Phys. Rev. {\bf D 22},
2227 (1980); M.~Magg and C.~Wetterich, Phys.~Lett.~B {\bf 94},~61 (1980);
R.~N.~Mohapatra and G.~Senjanovic, Phys. Rev. D{\bf 23}, 165 (1981);
G.~Lazarides, Q.~Shafi and C.~Wetterich, Nucl. Phys.~B{\bf 181}, 287 (1981)\,.

\bibitem{ma&sarkar_prl} E.~Ma and U.~Sarkar, Phys.\ Rev.\ Lett.
{\bf 80} (1998) 5716-5719\,.

\bibitem{type_II_group} S.~Antusch and S.~F.~King,
Phys.\ Lett.\ B {\bf 597}, 199 (2004); T.~Hambye and G.~Senjanovic, 
Phys.\ Lett.\ B{\bf 582}, 73 (2004); N.~Sahu and U.~Sarkar, 
Phys.\ Rev.\ D {\bf 74}, 093002 (2006); N.~Sahu and S.~Uma Sankar, 
Nucl.\ Phys.\ B {\bf 724}, 329 (2005); Phys.\ Rev.\ D {\bf 71}, 
013006 (2005); Pei-hong Gu and Xiao-jun Bi, Phys.\ Rev.\ D{\bf 70}, 
2004 (063511)\,.

\bibitem{susy_tev} L.~Boubekeur, T.~Hambye and G.~Senjanovic,
Phys.\ Rev.\ Lett.\  {\bf 93}, 111601 (2004);
[arXiv:hep-ph/0404038]; T.~Hambye, J.M.~Russel, S.M.~West,
JHEP {\bf 0407}, 070 (2004), [arXiv:hep-ph/0403183];
E.J.~Chun, [arXiv: hep-ph/0508050];

\bibitem{extra_tev} A partial list: N.~Sahu and U.A.~Yajnik,
Phys.\ Rev.\ D{\bf 71}, 2005 (023507), [arXiv:hep-ph/0410075];
 Phys.\ Lett.\ B {\bf 635} (2006) 11, [arXiv:hep-ph/0509285]; 
A.~Sarkar and U.~A.~Yajnik,
  Phys.\ Rev.\  D {\bf 76}, 025001 (2007)
  [arXiv:hep-ph/0703142]; 
S.~F.~King and T.~Yanagida,
  Prog.\ Theor.\ Phys.\  {\bf 114}, 1035 (2006)
  [arXiv:hep-ph/0411030]; 
A.~Abada, H.~Aissaoui and M.~Losada,
  Nucl.\ Phys.\  B {\bf 728}, 55 (2005)
  [arXiv:hep-ph/0409343].

\bibitem{leptogenesis-mnu-3N} A.~S.~Joshipura, E.~A.~Paschos and W.~Rodejohann,
  JHEP {\bf 0108} (2001) 029;
 S.~Pascoli, S.~T.~Petcov and W.~Rodejohann,
  Phys.\ Rev.\ D {\bf 68}, 093007 (2003);
J.~R.~Ellis and M.~Raidal,
  Nucl.\ Phys.\ B {\bf 643} (2002) 229;
G.~C.~Branco, T.~Morozumi, B.~M.~Nobre and M.~N.~Rebelo,
  Nucl.\ Phys.\ B {\bf 617} (2001) 475
  [arXiv:hep-ph/0107164].

\bibitem{leptogenesis-mnu-2N} K.~Bhattacharya, N.~Sahu, U.~Sarkar and 
S.K.~Singh, Phy.\ Rev.\ D\ {\bf 74}, 093001 (2006);
T.~Endoh, S.~Kaneko, S.~K.~Kang, T.~Morozumi and M.~Tanimoto,
  Phys.\ Rev.\ Lett.\  {\bf 89} (2002) 231601;
P.~H.~Frampton, S.~L.~Glashow and T.~Yanagida,
  Phys.\ Lett.\ B {\bf 548} (2002) 119\,.


\bibitem{singlet_lep} M.~Frigerio, T.~Hambye and E.~Ma,
JCAP {\bf 0609}, 009 (2006); N.~Sahu and U.~Sarkar,
  Phys.\ Rev.\  D {\bf 76}, 045014 (2007)
  [arXiv:hep-ph/0701062]. 

\bibitem{jm1}
  J.~McDonald, 
  Phys.\ Rev.\  D {\bf 50}, 3637 (1994)
  [arXiv:hep-ph/0702143].
  
\bibitem{jm2}
  J.~McDonald,
  Phys.\ Rev.\ Lett.\  {\bf 88}, 091304 (2002)
  [arXiv:hep-ph/0106249].

\bibitem{darkmatter}

  V.~Silveira and A.~Zee,
  Phys.\ Lett.\  B {\bf 161}, 136 (1985);
 C.~P.~Burgess, M.~Pospelov and T.~ter Veldhuis,
  Nucl.\ Phys.\  B {\bf 619}, 709 (2001)
  [arXiv:hep-ph/0011335];
V.~Barger, P.~Langacker and G.~Shaughnessy,
  Phys.\ Rev.\  D {\bf 75}, 055013 (2007)
  [arXiv:hep-ph/0611239];
H.~Sung Cheon, S.~K.~Kang and C.~S.~Kim,
  arXiv:0710.2416 [hep-ph].


\bibitem{singlet_resonant} M. Flanz, E.A. Paschos and U. Sarkar,
Phys.\ Lett. {\bf B 345}, 248 (1995); A.~Pilaftsis and
T.~E.~J.~Underwood, Nucl.\ Phys.\ B {\bf 692}, 303 (2004)\,.

\bibitem{triplet_resonant} G.~D'Ambrosio, T.~Hambye, A.~Hektor, 
M.~Raidal and A.~Rossi, Phys.\ Lett.\ B {\bf 604}, 199 (2004); 
E.~J.~Chun and S.~Scopel, Phys.\ Lett.\ B {\bf 636}, 278 (2006); 
[arXiv:hep-ph/0609259]\,. 



\bibitem{bento1}
  M.~C.~Bento, O.~Bertolami and R.~Rosenfeld,
  Phys.\ Lett.\  B {\bf 518}, 276 (2001)
  [arXiv:hep-ph/0103340].

\bibitem{bento2}
  M.~C.~Bento, O.~Bertolami, R.~Rosenfeld and L.~Teodoro,
  Phys.\ Rev.\  D {\bf 62}, 041302 (2000)
  [arXiv:astro-ph/0003350].

\bibitem{sidm}
  D.~N.~Spergel and P.~J.~Steinhardt,
  Phys.\ Rev.\ Lett.\  {\bf 84}, 3760 (2000)
  [arXiv:astro-ph/9909386].

\bibitem{boehm}
  C.~Boehm, D.~Hooper, J.~Silk, M.~Casse and J.~Paul,
  Phys.\ Rev.\ Lett.\  {\bf 92}, 101301 (2004)
  [arXiv:astro-ph/0309686].

\bibitem{posp}
  C.~Picciotto and M.~Pospelov,
  Phys.\ Lett.\  B {\bf 605}, 15 (2005)
  [arXiv:hep-ph/0402178].

\bibitem{beacom} J.~F.~Beacom and H.~Yuksel,
  Phys.\ Rev.\ Lett.\  {\bf 97}, 071102 (2006)
  [arXiv:astro-ph/0512411]; 
J.~F.~Beacom, N.~F.~Bell and G.~Bertone,
  Phys.\ Rev.\ Lett.\  {\bf 94}, 171301 (2005)
  [arXiv:astro-ph/0409403].

\bibitem{integral}
  P.~Jean {\it et al.},
  Astron.\ Astrophys.\  {\bf 407}, L55 (2003)
  [arXiv:astro-ph/0309484];
  J.~Knodlseder {\it et al.},
  Astron.\ Astrophys.\  {\bf 411}, L457 (2003)
  [arXiv:astro-ph/0309442].


\bibitem{mark}
  S.~W.~Randall, M.~Markevitch, D.~Clowe, A.~H.~Gonzalez and M.~Bradac,
  arXiv:0704.0261 [astro-ph];
  M.~Markevitch {\it et al.},
  Astrophys.\ J.\  {\bf 606}, 819 (2004)
  [arXiv:astro-ph/0309303].


\bibitem{collider_signature} G.~Barenboim, K.~Huitu, J.~Maalampi and M.~Raidal,
  Phys.\ Lett.\ B {\bf 394}, 132 (1997); K.~Huitu, J.~Maalampi, A.~Pietila
and M.~Raidal, Nucl.\ Phys.\ B {\bf 487}, 27 (1997); T.~Han, H.~E.~Logan,
B.~Mukhopadhyaya and R.~Srikanth, Phys.\ Rev.\ D {\bf 72}, 053007 (2005);
E.~Ma, M.~Raidal and U.~Sarkar, Nucl.\ Phys.\ B {\bf 615}, 313 (2001);
C.~Yue and S.~Zhao, [arXiv: hep-ph/0701017].

\bibitem{neutrinomass_pattern} A.~G.~Akeroyd and M.~Aoki, Phys.\ Rev.\ D {\bf 72}, 035011 (2005);
E.~J.~Chun, K.~Y.~Lee and S.~C.~Park, Phys.\ Lett.\ B {\bf 566}, 142 (2003).

\bibitem{newinterests} A.~G.~Akeroyd, M.~Aoki and H.~Sugiyama,
  arXiv:0712.4019 [hep-ph];
M.~Kadastik, M.~Raidal and L.~Rebane,
  arXiv:0712.3912 [hep-ph];
J.~Garayoa and T.~Schwetz,
  arXiv:0712.1453 [hep-ph].

\bibitem{eta_test} S.~Ando, J.~F.~Beacom, S.~Profumo and D.~Rainwater,
  arXiv:0711.2908 [hep-ph] and the references therein.









\end{thebibliography}
\end{document}